\documentclass{emulateapj}
\usepackage{apjfonts}
\usepackage{graphicx}
\usepackage{epsfig}
\usepackage{rotating}

\newcommand{\bc}{\begin{center}}
\newcommand{\ec}{\end{center}}
\newcommand{\be}{\begin{equation}}
\newcommand{\ee}{\end{equation}}
\newcommand{\ba}{\begin{eqnarray}}
\newcommand{\ea}{\end{eqnarray}}
\newcommand{\bt}{\begin{tabular}}
\newcommand{\et}{\end{tabular}}

\def\farcs{\hbox{$.\!\!^{\prime\prime}$}}
\def\farcm{\hbox{$.\!\!^{\prime}$}}

\def\farcs{\hbox{$.\!\!^{\prime\prime}$}}
\def\farcm{\hbox{$.\!\!^{\prime}$}}

\begin{document}

\submitted{
\today
}

\title{X-ray emission from PSR B1800-21, its wind nebula,
and similar systems}

\author{
 O.\ Kargaltsev, G.\ G.\ Pavlov, and G.\ P.\ Garmire}
\affil{The Pennsylvania State University, 525 Davey Lab, University
Park, PA 16802, USA} \email{pavlov@astro.psu.edu}

\begin{abstract}
We detected X-ray emission from the Vela-like pulsar B1800$-$21 and
resolved its synchrotron nebula with the {\sl Chandra} X-ray Observatory.
 The pulsar's flux is $F_{\rm psr} = (1.4\pm 0.2)\times 10^{-14}$
ergs cm$^{-2}$ s$^{-1}$ in the 1--6 keV band.
Its spectrum can be described by a two-component
power law (PL) + blackbody model,
suggesting a mixture of thermal
and magnetospheric emission.
For a plausible hydrogen column density
$n_{\rm H}=1.4\times10^{22}$ cm$^{-2}$,
the
PL component has a slope $\Gamma_{\rm psr}=1.4\pm 0.6$ and
a luminosity $L_{\rm psr}^{\rm nonth}
\approx 4\times 10^{31}(d/4\,{\rm kpc})^2$
ergs s$^{-1}$.
 The properties of the thermal component
($kT\sim 0.1$--0.3 keV,
$L_{\rm psr}^{\rm bol} \sim 10^{31}$--$10^{33}$ ergs s$^{-1}$)
are very poorly constrained because of
the strong
interstellar absorption.
The compact, $\approx 7''\times 4''$,
inner pulsar-wind nebula (PWN), elongated perpendicular to the
pulsar's proper motion, is immersed in a fainter asymmetric
emission. The observed flux of the PWN, including its fainter
component, is $F_{\rm pwn}=(5.5\pm 0.6)\times10^{-14}$ ergs cm$^{-2}$
s$^{-1}$ in the 1--8 keV band. The PWN spectrum can be described by
a PL model:
$\Gamma_{\rm pwn}\simeq1.6\pm 0.3$,
$L_{\rm pwn} \approx 1.6\times 10^{32} (d/4\,{\rm kpc})^2$ ergs
s$^{-1}$, for $n_{\rm H}=1.4\times10^{22}$ cm$^{-2}$.
The elongation of the inner PWN with respect to the direction
of the pulsar's proper motion
suggests
that its X-ray emission emerges from a torus
associated with the termination shock in the equatorial pulsar wind.
Such an interpretation provides additional support for the alignment
between the pulsar's velocity
and the spin axis found for several other pulsars.
The asymmetry in the fainter, more extended emission
could be attributed to nonuniform properties of the ambient medium.
A lack of any signs of
bow-shock morphology
suggests that the pulsar moves subsonically in high-pressure
interiors of a supernova remnant.
However, similar to a few other Vela-like pulsars,
no supernova remnant
is seen in the {\sl Chandra} image, possibly because its
soft X-ray emission is absorbed by the interstellar medium.
The inferred PWN-pulsar properties
(e.g., the PWN X-ray efficiency,
$L_{\rm pwn}/\dot{E}\sim 10^{-4}$;
the luminosity ratio, $L_{\rm pwn}/L_{\rm psr}^{\rm nonth}\approx 4$; the pulsar wind pressure at the termination shock,
$p_s\sim 10^{-9}$ ergs cm$^{-3}$)  are very similar to those
of other subsonically moving Vela-like objects detected with {\sl Chandra\/}
($L_{\rm pwn}/\dot{E}\sim 10^{-4.5}$--$10^{-3.5}$,
$L_{\rm pwn}/L_{\rm psr}^{\rm nonth} \sim 5$,
$p_s\sim 10^{-10}$--$10^{-8}$ ergs cm$^{-1}$).

\end{abstract}
\section{Introduction}

Observations with the {\sl Chandra} X-ray Observatory have shown
that many young pulsars ($\tau\lesssim 30$  kyrs) power X-ray
nebulae
 (e.g., Kaspi et al.\ 2004; Gaensler \& Slane 2006, and
references therein). The observed X-ray emission is
produced by relativistic particles gyrating in the magnetic
field downstream of the termination shock
 in the pulsar
wind (Kennel \& Coroniti 1994; Arons 2004).
 The innermost parts of many X-ray pulsar-wind nebulae (PWNe) show
axisymmetric morphologies,
including toroidal structures and jets along
the pulsar's spin axis.
However, even pulsars with very similar
spin-down properties (such as
the period, $P$, period derivative, ${\dot{P}}$,
spin-down power, ${\dot{E}}=4\pi I
\dot{P}P^{-3}$, and
spin-down age, $\tau=P/2\dot{P}$)
can produce PWNe of quite different
shapes and sizes.
These
differences can occur for a number of reasons. For instance,
upstream of the termination shock (i.e., closer to the pulsar) the
properties of the wind should be sensitive to the
 angle between the pulsars's magnetic and spin axes
and the wind magnetization.
These factors could affect the strength of the termination shock
and, consequently, the temperature and radiation efficiency of the
post-shock flow.
On the other hand,
the  properties of the post-shock flow
should also depend on density and temperature of the ambient
medium. For instance, these parameters may vary significantly for
PWNe residing inside supernova remnants (SNRs) of different
ages.
 Finally, the direction
of the kick acquired by the neutron
star (NS) at birth with respect to
its spin axis, and the NS velocity
relative to the ambient medium
also affect the PWN
appearance.
For instance, if the pulsar moves with a high velocity, the
ram pressure caused by its motion can exceed the ambient gas
pressure, resulting in a
 bow-shock PWN,
such as ``the Mouse'' PWN around PSR J1747$-$2958 (Gaensler et al.\ 2004)
  and ``the Duck'' PWN around PSR B1757$-$24 (Kaspi et al.\ 2001a).
 Studying X-ray bright, nearby PWNe helps to
disentangle various effects, understand their impact on the PWN
structure, and
probe the properties of the ambient medium.

The
well-known nearby ($d\approx 300$ pc)
Vela pulsar and its X-ray PWN (Pavlov et al.\ 2001ab,
2003; Helfand et al.\ 2001) have become an archetype for
young ($\tau\sim 10$--30 kyrs) and energetic
($\dot{E}\sim 10^{36}$--$10^{37}$
ergs s$^{-1}$) pulsars
powering X-ray PWNe. These pulsars are also interesting because
 at this age emission from the hot NS surface
becomes observable in X-rays
 as a thermal ``hump'' on top of
the flat non-thermal spectrum (e.g., Kargaltsev \& Pavlov 2006).

Based on its age, $\tau =16$ kyr, and spin-down power,
$\dot{E}=2.2\times 10^{36}$ ergs s$^{-1}$, the
radio pulsar B1800--21 (hereafter B1800)
is similar to the Vela pulsar, but it is more distant
(the dispersion measure distance is $d=3.8\pm 0.4$ kpc,
according to Cordes \& Lazio 2002).
Its spin-down
flux, $\dot{E}/4\pi d^{2}=1.2 \times 10^{-9}d_{4}^{-2}$ ergs s$^{-1}$
cm$^{-2}$, where
$d_{4}\equiv
d/(4\, {\rm kpc})$,
places it among the top 20 pulsars ranked by this parameter.
The pulsar is projected
near the western boundary of the radio SNR
  G8.7$-$0.1 (W30),
surrounded by
HII regions and molecular gas (Odegard 1986, and references therein).
 However, the association between B1800 and G8.7$-$0.1 has been
considered doubtful because it would require a very large
velocity of the pulsar if it was born near the apparent SNR center
(Frail et al.\ 1994).

Being young and energetic, B1800
should be surrounded by a PWN, and both the pulsar and the PWN
should be detectable in X-rays.
Based on a 10 ks {\sl ROSAT} PSPC observation, Finley \& \"Ogelman
(1994) have reported a faint X-ray source (PSPC count rate of
$1.5\pm0.5$ counts ks$^{-1}$) located near the radio pulsar position
 and attributed this emission
to B1800.
The low count rate and insufficient angular resolution
of {\sl ROSAT} PSPC
did not allow
 Finley \& \"Ogelman (1994) to resolve a compact PWN.
They, however, detected diffuse X-ray emission
$\sim 30'$ northeast of the pulsar and attributed it to the
G8.7$-$0.1 SNR. These authors defend the association between the pulsar and
G8.7$-$0.1 by suggesting that the SN explosion, which produced
B1800 and G8.7$-$0.1,
occurred in or near a molecular cloud, very close to the current position
of the pulsar, and the SNR have been ``blown out'' eastward into a low-density
interstellar medium (ISM).
However, a recent proper motion measurement (Brisken et al.\ 2006)
has shown that the pulsar
was born outside the currently seen SNR, and it moves more nearly
toward the center of
G8.7$-$0.1 rather than away from it,
which makes their association very unlikely.
This measurement also essentially rules out the
association between the pulsar and the newly discovered
SNR candidate
G8.31$-$0.09
(Brogan et
al.\ 2006).

B1800 is of particular interest because it is located in the vicinity
of the TeV $\gamma$-ray source HESS J1804$-$216 (Aharonian et al.\ 2006),
and it may supply ultrarelativistic electrons that generate the
TeV radiation by upscattering the photons of the cosmic microwave background.
We will discuss the possible connection between B1800 and HESS J1804$-$216,
as well as other candidate X-ray counterparts of the TeV source,
in a separate paper.

In this paper, we describe the results of a {\sl Chandra}
observation of PSR B1800--21 and its synchrotron nebula. The
details of the observation and the data analysis are presented in
\S2. We compare the X-ray properties of the B1800 pulsar and its
PWN with those of other Vela-like pulsars-PWNe
and discuss implications of our findings in \S3.
Our main results are summarized in \S4.

\section{Observations and Data Analysis}

We observed B1800 with the Advanced CCD Imaging Spectrometer (ACIS)
on board {\sl Chandra} on 2005 May 4.
 The useful scientific exposure time was 30,236 s. The
observation was carried out in Faint mode, and the pulsar was imaged
on S3 chip, $\approx7\farcs{5}$ from the aim point. The detector was
operated in Full Frame mode, which provides time resolution of 3.2
seconds. The data were reduced using the Chandra Interactive
Analysis of Observations (CIAO) software (ver.\ 3.2.1; CALDB ver.\
3.0.3).

\begin{figure*}[t]
 \centering
\includegraphics[width=7.2in,angle=0]{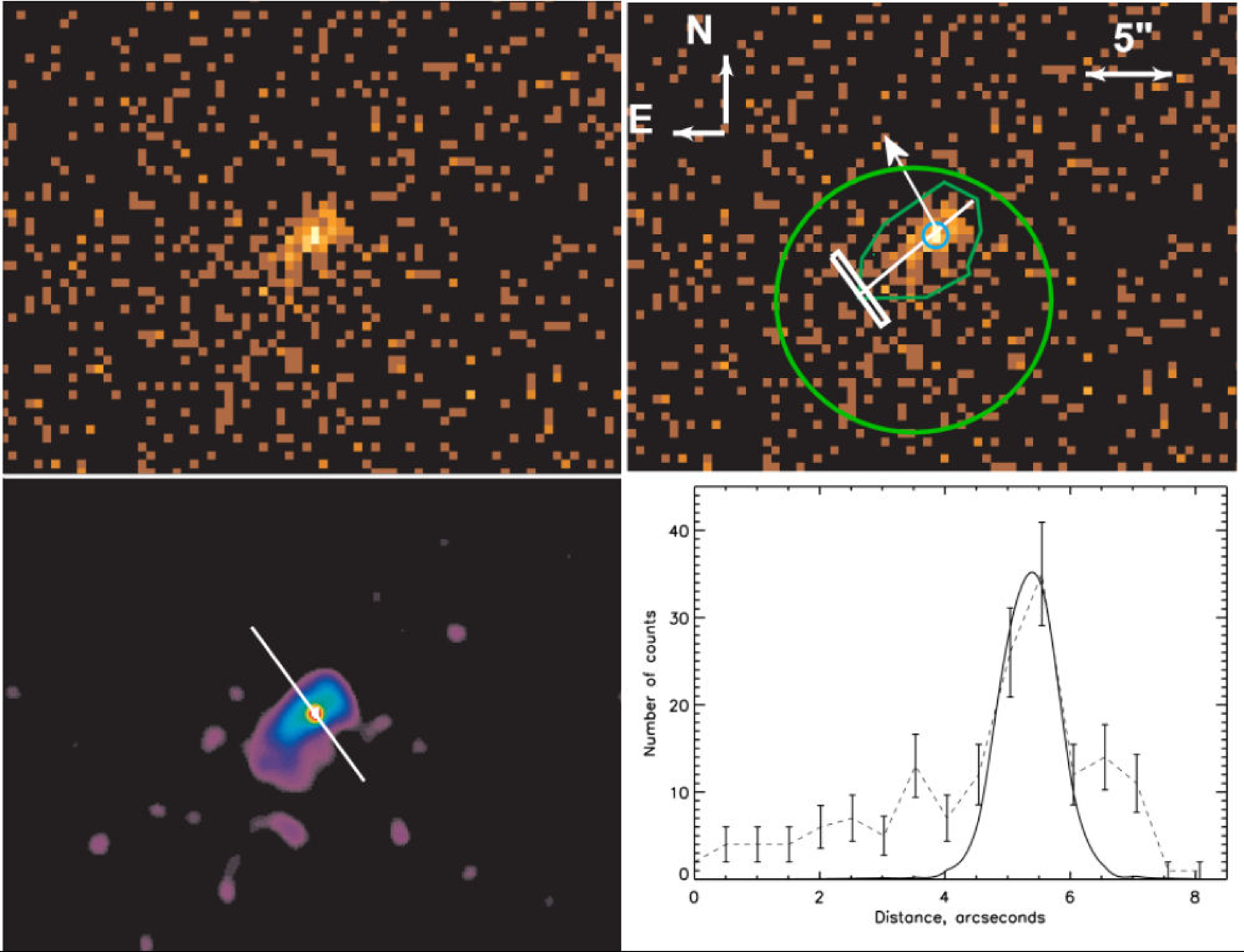}
 \caption{ {\em Top left:} ACIS-S3 image
 of B1800 and its PWN. {\em Top right:}
Extraction regions used for the spectral and image
analysis of the source components
(green and white lines; see
 text). The white arrow shows the direction of the pulsar proper motion
(Brisken et al.\
 2006);
its length corresponds to
the projected distance
traveled by the pulsar
in 300 years.
 {\em Bottom left:} The smoothed sub-pixel (0.25 of the original ACIS pixel)
resolution image obtained by removing the pipeline
pixel-randomization and subsequently applying the sub-pixel
resolution tool, based on analyzing the charge distribution produced
by an X-ray event (Tsunemi et al.\ 2001; Mori et al.\ 2001)
and maximum-likelihood
deconvolution procedure (Richardson 1972; Lucy 1974). The
straight line shows
the rotation axis of the putative torus (see text). {\em
Bottom right:} Intensity distribution
 along the
line perpendicular to the rotation axis
measured in $0\farcs5\times5''$ box
moving along the line
 shown in the top right panel. The solid line shows
 the simulated one-dimensional point-spread function of the {\sl Chandra} ACIS.
}
\end{figure*}

\begin{figure}[t]
 \centering
\includegraphics[width=3.2in,angle=0]{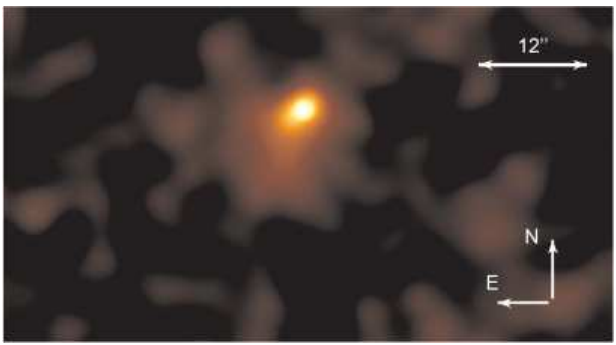}
 \caption{ Adaptively smoothed
 ACIS-S3 image of B1800 and its PWN. The brightness and smoothing scales are chosen
 to show the fainter, more extended emission from the PWN.}
\end{figure}

\subsection{Image}

Figure 1 shows the ACIS-S3 image of the region around B1800. An
extended X-ray source is clearly seen in the image around
R.A.$=18^{\rm h}03^{\rm m}51\fs432$, decl.$=-21^{\circ}37'
07\farcs45$ (these are the coordinates of the center of the
brightest pixel).
The difference of $0\farcs3$ between this postion
and the radio
position\footnote{
Note that the radio timing position of B1800,
 R.A.$=18^{\rm h}03^{\rm
m}51\fs333(8)$, decl.$=-21^{\circ}36' 27''(3)$
(Manchester et al.\ 2005), differs by $\approx40''$ from the Brisken et
al.\ (2006) position and the position of the X-ray source. No X-ray
source is seen at or near the
radio timing position. This means that the error
of the radio timing position is strongly underestimated.}
of B1800 (Brisken et al.\ 2006)
is smaller than
the error in absolute {\sl Chandra} astrometry (0\farcs6 at the 90\%
confidence level).
 The close match between the X-ray and radio positions and
  the extended morphology of the observed
X-ray emission allow us to conclude that we detected the X-ray emission
from B1800 and its PWN.

The brighter, compact ($\approx 7''\times4''$, i.e.
$0.14\times 0.08$ pc$^2$ at $d=4$ kpc)
PWN component
is elongated in the
southeast-northwest direction,
roughly perpendicular to the proper motion direction (see Fig.\ 1).
 The compact component
(the inner PWN)
is brighter
northwest of the pulsar
but more extended toward
southeast.
There is also evidence for a more extended,
fainter
emission component (the outer PWN), $\sim 12''$ in size, mostly concentrated
  southward
 of the compact bright component
(Fig.\ 2). The bottom right panel in Figure 1 shows
the one-dimensional distribution of counts along the
southeast-northwest direction. The data points with error bars
(connected by the dashed line) are obtained by integrating counts
within $10\times1$ pixels ($5''\times 0\farcs5$) rectangular
apertures (shown in the top right panel of Fig.\ 1) moving along the
southeast-northwest direction with a 0\farcs5 step.
   The solid line
shows the count distribution
obtained in a similar way but for a point source simulated with
    MARX\footnote{MARX (Model of AXAF
Response to X-rays) is a suite of programs designed to enable the
user to simulate the on-orbit performance of the {\sl Chandra}
satellite, including ray-trace simulation. See
http://space.mit.edu/ASC/MARX/}. The difference between the dashed
 and solid lines
represents the extended emission from the PWN. The one-dimensional
profile and the PWN images suggest a compact ($\sim4''-5''$ in
diameter),
roughly symmetric
torus around the pulsar (the blue region in the bottom left panel of Fig.\ 1),
with some extension toward southeast.
The position angle
(PA) of the torus symmetry axis is about $48^\circ$--$50^{\circ}$
east of north,
close to the proper motion PA of $38\fdg 1\pm 6\fdg 3$
(Brisken et al.\ 2006).

We also attempted to search for signatures of an SNR around the pulsar-PWN
complex. A direct visual inspection of the ACIS image did not show
clear signatures of large-scale diffuse emission.
We applied the exposure map correction and smoothed the image with
various scales, but failed to find statistically significant
deviations from a uniform brightness distribution
in these images.
To estimate an upper limit on the SNR emission,
we measured the
count rate from the entire S3 chip (with all identifiable point
sources removed). The count rate,
$0.596\pm0.004$ counts s$^{-1}$
  in the 0.5--7 keV band,
exceeds the nominal S3 background of 0.32 counts s$^{-1}$
({\sl Chandra} Proposers' Observatory
Guide\footnote{See http://asc.harvard.edu/proposer/POG/index.html}, v.8, \S6.15.2),
which could be caused by an elevated particle background,
diffuse X-ray background, or SNR emission. Since we see no trace of an SNR,
we consider the difference, 0.28 counts s$^{-1}$,
 as an upper limit on the SNR count rate in the 70 arcmin$^2$ of the chip
area, which corresponds the average surface brightness limit of
4 counts ks$^{-1}$ arcmin$^{-2}$.

\subsection{Spectral analysis}

\subsubsection{PWN spectrum}

The PWN spectra were extracted from two regions shown in Figure 1
(top right panel). The smaller (polygon) region of 32.7 arcsec$^2$
area encompasses the brighter compact PWN, while the larger
(circular) region of 176 arcsec$^2$ area includes a fainter
component, which is more prominent south of the pulsar. From these
regions we excluded the circular region
of $1\farcs46$ radius
 centered on the brightest pixel to avoid contamination
of the PWN spectrum by the pulsar. The background was
 measured in the
$20''<r<25''$ annulus
centered on the source
(45 counts in the 706 arcsec$^2$ area, in the 0.3--8 keV band).
 The total numbers of counts
extracted from the smaller and larger PWN regions are 58 and 117, of
which 96.4\% and 90.3 \% are expected to come from the source, which
gives $55.9\pm 7.6$  and $105.6\pm 11.6$ PWN counts in the two
regions. (The errors here and below are at the 68\% confidence level.)
The observed PWN fluxes are $F_{\rm pwn} =
(5.5\pm0.6$) and ($2.7\pm0.4$) $\times 10^{-14}$ ergs s$^{-1}$
cm$^{-2}$ for the larger and smaller extraction regions,
respectively, in the 1--8 keV band.

For each of the two spectra, we group the counts into six
spectral bins, with comparable numbers of counts per bin.
First, we fit the PWN spectra for each of the two regions with the
absorbed PL model, allowing the hydrogen column density, $n_{\rm H}$,
to vary.
 These fits
result in
 spectral slopes $\Gamma_{\rm pwn}\approx1.5$--1.9
 and
$n_{\rm H,22}\equiv n_{\rm H}/10^{22}\, {\rm cm}^{-2} \approx1.2$--2.0
 (see Fig.\ 3).
 We see no indication for spectral
softening (expected due to synchrotron burn-off in PWNe with
high radiation efficiency)
 in the spectrum extracted from the larger region compared
to the spectrum of the bright inner PWN. To
test this, we
also fit the spectra with the PL model in which the hydrogen
column density is
fixed at $n_{\rm H,22}=1.38$ (the best-fit value for the larger
region) and found that the
virtually the same photon indices:
$\Gamma_{\rm pwn}=1.58\pm 0.25$
 and $1.55\pm 0.29$, for the
larger and smaller region, respectively.
 The apparent lack of
spectral softening could
 be attributed to the small number of counts
collected from the fainter outer PWN component.

The best-fit value of $n_{\rm H,22}\approx 1.4$ estimated from
the B1800 PWN fit approximately coincides with the
total Galactic HI column density in this direction (Dickey \&
Lockman 1990). It does not contradict to the adopted distance
of $\approx$4 kpc to B1800 because the $n_{\rm H}$ deduced from
an X-ray spectrum under the assumption of standard element
abundances generally exceeds the $n_{\rm HI}$
measured from 21 cm observations
by a factor of 1.5--3
(e.g., Baumgartner \&
Mushotzky 2005).
On the other hand,
given the B1800's dispersion measure, ${\rm DM}=234$ cm$^{-3}$ pc
(i.e., the electron column density $n_e = 7.22\times 10^{20}$ cm$^{-2}$),
the $n_{\rm H,22}$ value of 1.4 corresponds to the ISM ionization
degree
$n_e/n_{\rm H} \approx 5\%$, below the usually assumed
value of 10\%, which corresponds to $n_{\rm H,22}= 0.72$.
If the hydrogen column density is fixed at $n_{\rm H,22}=0.72$,
the PL fit results in substantially harder spectra,
  $\Gamma_{\rm pwn}\approx 1.0$ for both extraction regions (see
Table 1 and Figs.\ 3 and 4).

\begin{figure}
 \centering
\includegraphics[width=2.5in,angle=90]{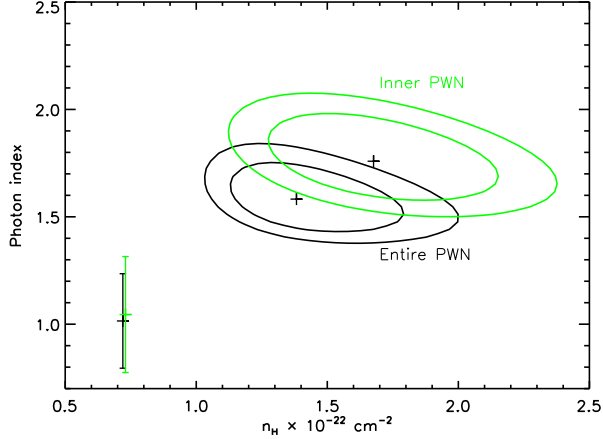}
\caption{Confidence contours (68\% and 90\%)
 in the $n_{\rm
H}$--$\Gamma$ plane for the PL fit to the PWN spectrum. The error
bars show the photon indexes for the fixed $n_{\rm H, 22}=0.72$.
Black and green
contours and error bars correspond to the
larger and smaller PWN extraction regions, respectively (see text and Fig.\ 2).
}
\end{figure}

\begin{figure}
 \centering
\includegraphics[width=2.5in,angle=90]{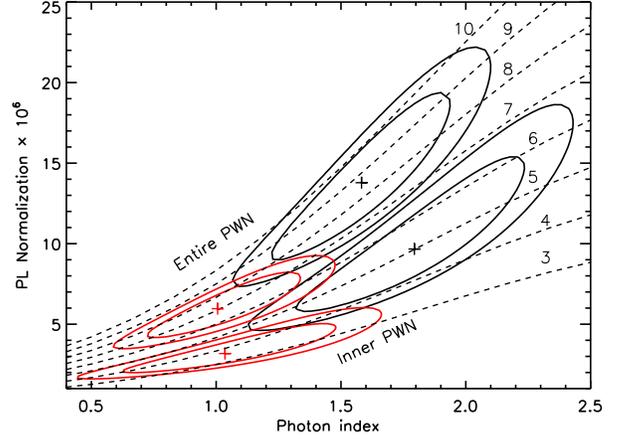}
\caption{
 Confidence contours
(68\% and 90\%) for the PL fit to the PWN spectra with
$n_{\rm H,22}=1.38$ (black)
and
0.72 (red). The upper contours correspond to the larger extraction
region.
 The PL normalization
is in units of $10^{-6}$ photons cm$^{-2}$ s$^{-1}$
keV$^{-1}$ at 1 keV. The dashed curves are the loci
 of constant unabsorbed flux in the 0.5--8 keV band;
the flux values near the curves are
in units of 10$^{-14}$ ergs cm$^{-2}$ s$^{-1}$.
}
\end{figure}

\begin{figure}
  \centering
\includegraphics[width=2.5in,angle=90]{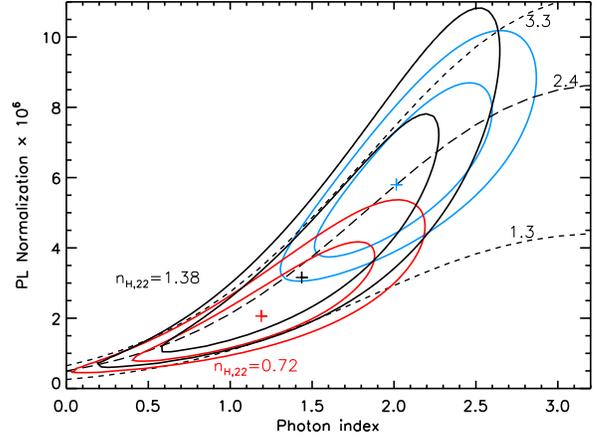}
\caption{
 Confidence contours (68\% and 90\%) for PL spectral parameters
for various fits to the pulsar spectrum:  PL fit
with $n_{\rm H,22}=0.72$ (blue), PL+BB fit with $n_{\rm H,22}=1.38$
(black),
 and PL+BB fit with
$n_{\rm H,22}=0.72$ (red).
 The PL normalization
is in
units of $10^{-6}$ photons cm$^{-2}$ s$^{-1}$ keV$^{-1}$ at 1 keV.
The dashed curves are the lines of constant unabsorbed flux
in the 0.5--8 keV band (the flux values are in units of
$10^{-14}$ ergs cm$^{-2}$ s$^{-1}$).
}
\end{figure}

The unabsorbed flux
and luminosity are less sensitive to the interstellar absorption
than the spectral parameters.
The
isotropic PWN
luminosity
is $L_{\rm pwn}\equiv 4\pi d^2 F_{\rm pwn}^{\rm unabs} \approx$
1--2 and 0.6--1 $\times 10^{32}d_{4}^{2}$ ergs s$^{-1}$ in the
0.5--8 keV band, for the whole and inner PWN, respectively
(see Table 1 and Fig.\ 4).
The former value
is a fraction of $\approx (0.5$--$1)\times 10^{-4} d_4^2$ of the pulsar's
spin-down power $\dot{E}$.

\subsubsection{Pulsar spectrum}

To minimize contamination by the PWN,
the pulsar spectrum was extracted from a small circular aperture
with the radius of 1.5 ACIS pixels ($\simeq 0\farcs 74$, 85\%
encircled energy radius).
 The
total number of counts within the source region is 51, of which less
than 15\% is expected to come from the PWN (based on the PSF
simulation). To account for the PWN contribution to the pulsar
 spectrum, the background spectrum was extracted from the $8$ arcsec$^{2}$ region
encompassing the brightest part of the inner PWN (with the pulsar
being excluded). The pulsar's absorbed flux
is $F_{\rm psr}=(1.4\pm0.2) \times10^{-14}$ ergs cm$^{-2}$ s$^{-1}$ in the 1--6
keV band (aperture corrected and background subtracted).
 To obtain constrained fits with the small number of counts available, we are forced to freeze the
hydrogen column density. First, we fix it at $n_{\rm H,22}=1.38$,
obtained above from the PL fit to the PWN spectrum.
With this $n_{\rm H}$, the single-component PL
and black-body (BB) models fail to fit the pulsar spectrum.
A two-component BB+PL model
provides a good fit, but the fitting parameters are poorly
constrained because of the small number of photons detected (Fig.\
5). The slope of the PL component is $\Gamma_{\rm psr}= 1.4\pm 0.6$,
and its unabsorbed luminosity is
$L_{\rm psr}^{\rm nonth}=(4.4\pm1.1) \times 10^{31} d_4^2$ ergs s$^{-1}
\approx2\times10^{-5}\dot{E} d_4^2$, in the 0.5$-$8 keV band.
The temperature and the
projected area
 of the BB component are
strongly correlated (see Fig.\ 6), which results in very large
uncertainties for these parameters. The
 best-fit temperature
is $T\approx1.6$ MK,
  while the projected emitting
  area,
 $\mathcal{A}\sim 2\times 10^{7} d_4^2$ m$^2$,
is smaller than that of the NS surface ($\pi R^{2}\sim3\times
10^{8}$ m$^{2}$),
but larger than the
conventional polar cap area $A_{\rm pc}=2\pi^2R^3/cP\approx 5\times10^5$
 m$^{2}$. The corresponding bolometric luminosity,
$L_{\rm psr}^{\rm bol} \equiv 4{\mathcal{A}}\sigma T^4 \sim 3\times 10^{32}
d_4^2$ ergs s$^{-1}$.

 If the hydrogen
column is fixed at the
value estimated from the dispersion measure at 10\% ISM ionization,
$n_{\rm H,22}=0.72$, then a single PL model fits the spectrum and
gives $\Gamma_{\rm psr} = 2.0^{+0.4}_{-0.3}$ and the unabsorbed
luminosity $L_{\rm psr}^{\rm nonth}\approx(4.9\pm0.8)\times 10^{31} d_4^2$
ergs s$^{-1}$, in the 0.5--8 keV band.
 Once again, the BB
model does not fit the spectrum while
 the
two-component PL+BB model
 provides an acceptable fit.
In comparison with the BB+PL fit with the larger $n_{\rm
H,22}=1.38$, the BB component shows a higher temperature ($T\approx
2.6$ MK) and a smaller (but even more uncertain) emitting area
($\mathcal{A}\sim 2\times 10^{5} d_4^2$ m$^{2}$) while the PL
component has $\Gamma_{\rm psr}\approx 1.2$.
The fitting parameters of the BB component  are
even less constrained than those at $n_{\rm H,22} =1.38$
(see Fig.\ 6). For the non-thermal PL component, the confidence
contours for fitting parameters strongly overlap with those
calculated for $n_{\rm H,22}=1.38$
 (see Fig.\ 5).
 The
luminosity of the non-thermal component,
$L_{\rm psr}^{\rm nonth}\approx
(3.7\pm 0.9)\times 10^{31} d_4^2$ ergs s$^{-1}$ in 0.5$-$8 keV,
is close to that in the $n_{H,22}=1.38$ case, while
the bolometric luminosity of the thermal component, $L_{\rm psr}^{\rm
bol}
\sim 2\times 10^{31} d_4^2$ ergs s$^{-1}$,
 is a factor of
15
 lower.

\begin{figure}[]
 \centering
\includegraphics[width=2.5in,angle=90]{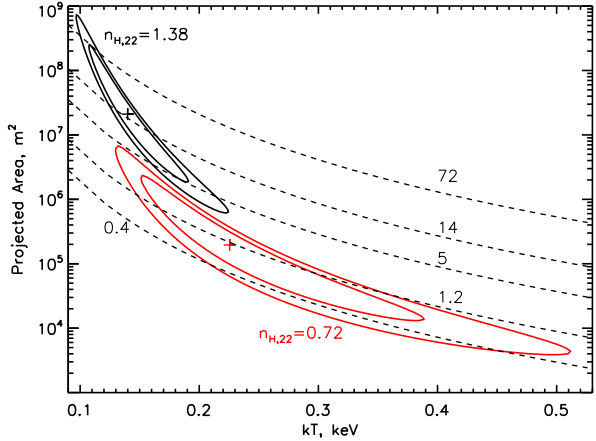}
\caption{ Confidence contours (68\% and 90\%) for the BB component
of the BB+PL fit to the pulsar's spectrum, for $n_{\rm H,22}=1.38$
(black), and $n_{\rm H,22}=0.72$ (red).
 The BB normalization (vertical axis) is the
projected emitting area in units of m$^2$, assuming the distance of
4 kpc. The lines of constant bolometric flux (in units of 10$^{-14}$
ergs cm$^{-2}$ s$^{-1}$) are plotted as dashed lines.
 }
\end{figure}

\begin{table}[]
\caption[]{PL fits to the PWN spectrum} \vspace{-0.5cm}
\begin{center}
\begin{tabular}{ccccccc}
\tableline\tableline Model & $n_{\rm H,22}$\tablenotemark{a}   &
$\mathcal{N}$\tablenotemark{b}  &
$\Gamma$ & ($C$ or $\chi^2$)\tablenotemark{c}/dof   & $L_{\rm X}$\tablenotemark{d} \\
\tableline

  Entire PWN           &       $1.38$     &
$13.8^{+3.5}_{-3.5}$       &
$1.58^{+0.25}_{-0.25}$          &  $0.93/3$  & $1.65_{-0.17}^{+0.16} $  \\
  Entire PWN           &       $0.72$                     &
$6.0^{+1.5}_{-1.3}$ &       $1.01_{-0.18}^{+0.23}$   &  $2.16/4$  & $1.35_{-0.18}^{+0.11} $  \\
  Inner  PWN         &       $1.38$     &
$9.7^{+3.7}_{-2.7}$ &       $1.79_{-0.32}^{+0.29}$   &  $1.34/3$   & $0.96_{-0.13}^{+0.16}$   \\
  Inner  PWN         &       $0.72$                     &
$3.3^{+1.2}_{-0.8}$  &       $1.04_{-0.26}^{+0.29}$   &  $1.34/4$   & $0.70_{-0.10}^{+0.11}$    \\

      \tableline
\end{tabular}
\end{center}
\tablenotetext{a}{The values of hydrogen column density are frozen in the fits.}
\tablenotetext{b}{Spectral flux in units of $10^{-6}$ photons
cm$^{-2}$ s$^{-1}$ keV$^{-1}$ at 1 keV.}
\tablenotetext{c}{We use
the C statistic (Cash 1979) for the inner PWN and $\chi^2$ statistic
for
entire PWN, which has a larger number of counts.}
\tablenotetext{d}{Unabsorbed isotropic luminosity in the 0.5--8 keV band in
units of $10^{32} d_4^2$ ergs s$^{-1}$.
}
\end{table}

\section{Discussion.}

The B1800 PWN+pulsar flux measured by {\sl Chandra} is consistent with the
{\sl ROSAT}\  PSPC count rate reported by Finley \& \"{O}gelman
(1994). However, {\sl ROSAT} was unable to resolve the PWN emission
from that of the pulsar. The superior angular resolution of {\sl
Chandra} has allowed us
to resolve the shape of the
PWN, disentangle the point source and the extended emission
components, and study their properties separately. Below we
discuss these properties in more detail and compare the results
on B1800 with those on other Vela-like pulsars.

\subsection{Energetics and spectra of B1800 and other Vela-like
pulsars and PWNe}

Since the pulsar wind energetics is supplied by the loss
of the pulsar spin energy,
it is natural to expect that the PWN luminosity is correlated
with the pulsar spin-down power, $\dot{E}$.
Such a correlation
was first noticed in X-ray PWN observations
with the {\sl Einstein} observatory (Seward \& Wang 1988).
On the other hand, as the wind properties can depend on,
e.g., pulsar's magnetic field and the angle between its magnetic
and spin axes, and the PWN properties depend on pulsar's velocity and
pressure of ambient medium,
we should not expect that the X-ray PWN efficiency,
$\eta_{\rm pwn}
\equiv L_{\rm pwn}/\dot{E}$, is the same for all PWNe. To examine the
$L_{\rm pwn}$-$\dot{E}$ correlation for
similar PSR/PWN systems ($\tau = 10$--30 kyrs, $\dot{E}=10^{36}$--$10^{37}$
ergs s$^{-1}$)\footnote{Pulsars with such $\tau$ and $\dot{E}$ are
traditionally dubbed Vela-like pulsars.},
we have estimated the X-ray PWN
luminosities in the 0.5--8 keV band
 for 9 such objects (in addition to B1800) using archival
{\sl Chandra} data (see Table 2).
Figure 7 demonstrates
that although there is a positive correlation
between $L_{\rm pwn}$ and $\dot{E}$, it is far from linear and shows
a large scatter. In particular, the PWN efficiency spans over
2.5 orders of magnitude, $10^{-4.5}\lesssim \eta_{\rm pwn} \lesssim 10^{-2}$
 in the relatively narrow $\dot{E}$ range,
$(2$--$7)\times 10^{36}$ ergs s$^{-1}$ (the B1800 PWN luminosity
and efficiency are among the lowest in this sample).
Such a scatter can hardly be explained by poorly known distances
to some of the objects or by the bias caused by a possible underestimation
of $L_{\rm pwn}$ for distant PWNe, where we can miss a  component with
lower surface brightness.
Interestingly, the two bow-shock PWNe
with prominent tails,
the Mouse
 and the Duck, show efficiencies higher than most of
the other PWNe in the sample,
perhaps because the wind is channeled into a smaller volume
produces a brighter X-ray image.
Another very bright
and efficient PWN is that around PSR J1811--1925, which is located in
the young SNR G11.2--0.3, possibly associated with the historical SN of 386 AD
(Kaspi et al.\ 2001b). If this association is correct, the true pulsar age is a factor
of 14 smaller than the
spin-down age, so that PSR J1811--1925
is much younger than the other pulsars in the sample, which may
explain why its PWN is so different. Excluding these three PWNe from
the sample substantially narrows the ranges of PWN
luminosities
and efficiencies:
\be
L_{\rm pwn}\sim 10^{32} - 10^{33}\,\, {\rm ergs\,\, s}^{-1}, \quad
\eta_{\rm pwn} \sim 10^{-4.5} - 10^{-3.5}\, .
\ee

\begin{figure}[]
 \centering
\includegraphics[width=2.7in,angle=90]{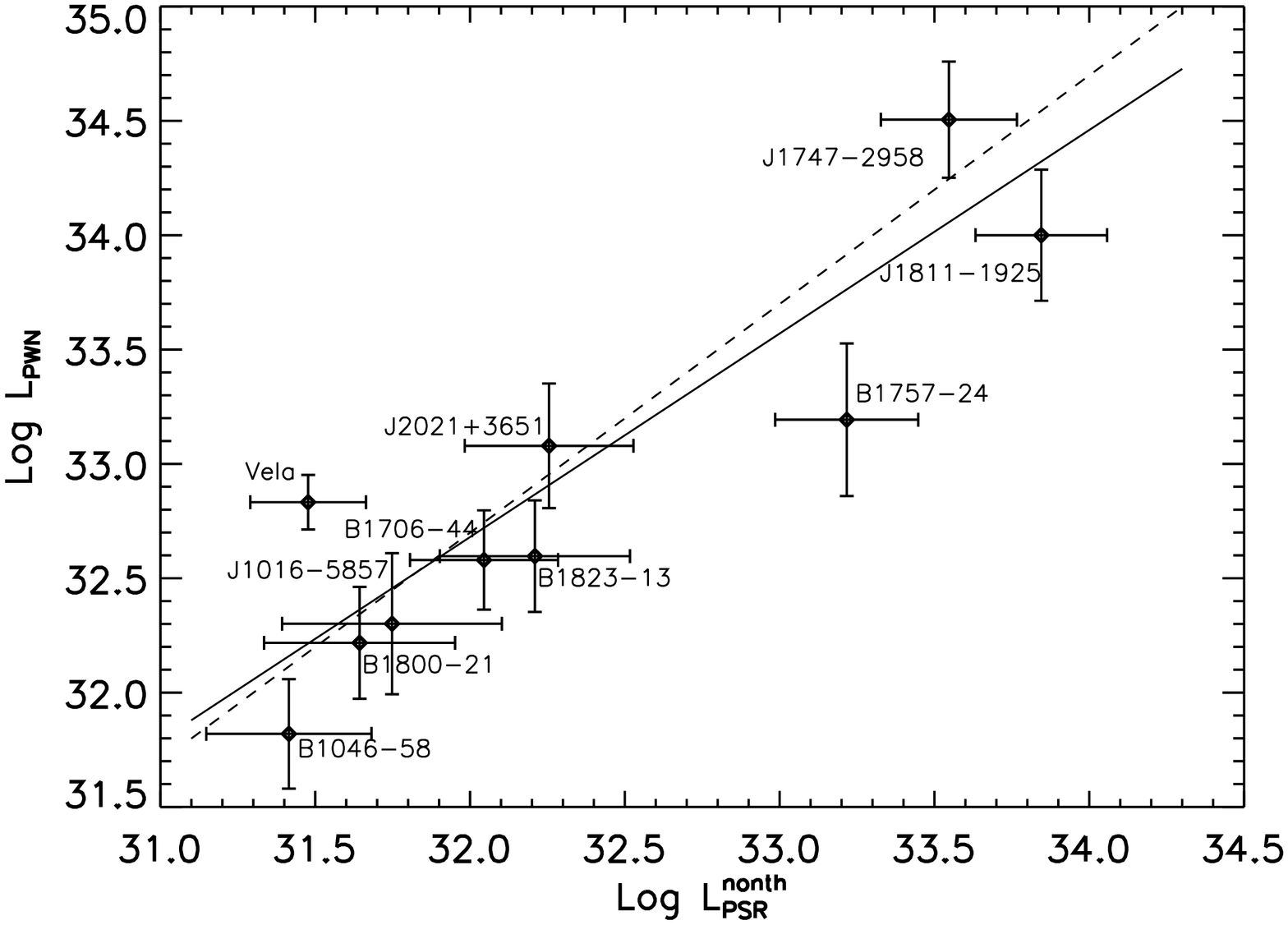}
\includegraphics[width=2.7in,angle=90]{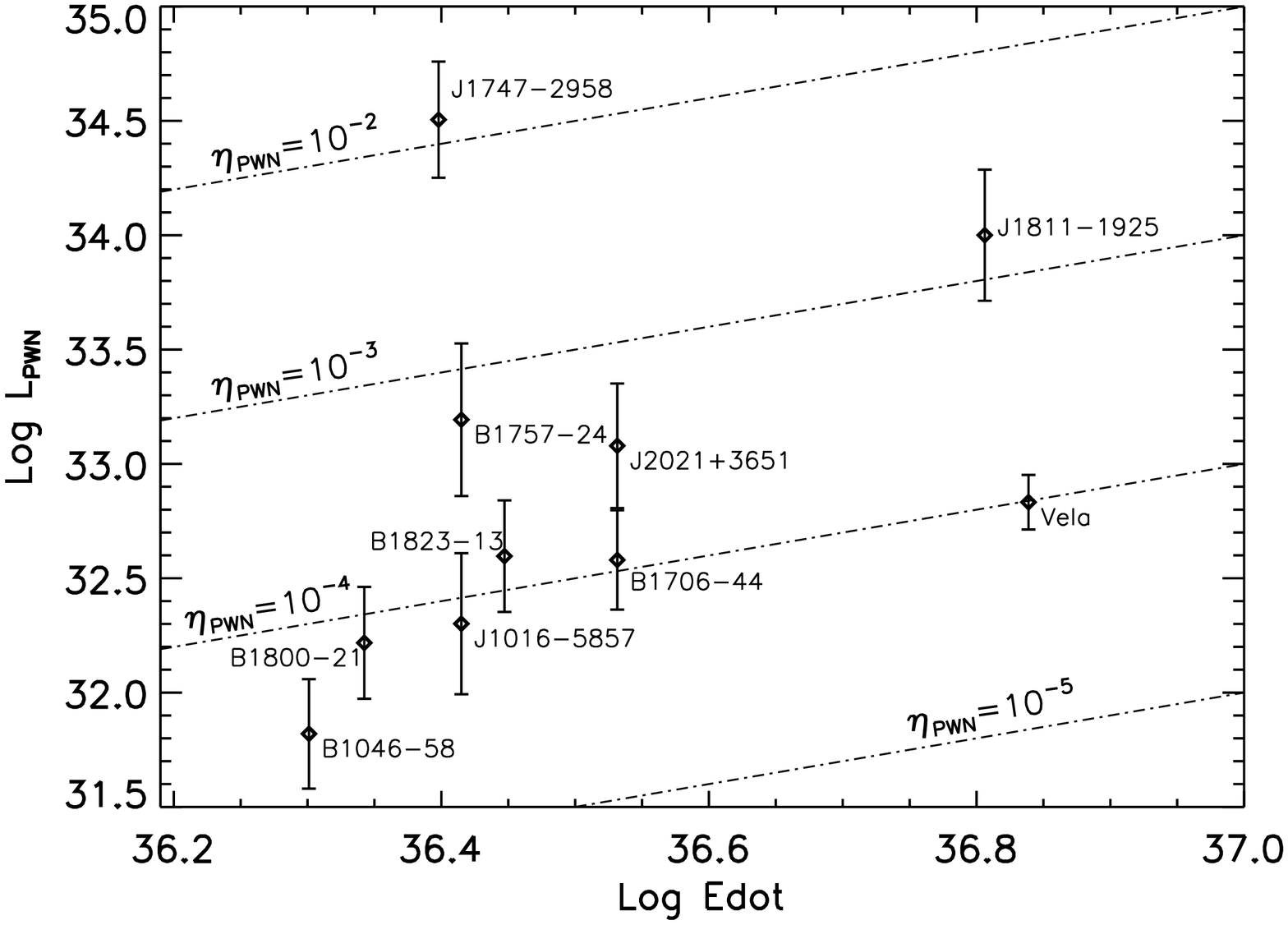}
\caption{
PWN luminosity versus pulsar non-thermal luminosity
({\em top}) and
spin-down power ({\em bottom}) for
10 Vela-like pulsars observed with {\sl Chandra}.
The luminosities are estimated for the 0.5--8 keV band.
The solid
and dashed
 lines in the {\em top} panel
correspond to
$\log L_{\rm pwn}=0.89\log L_{\rm psr}^{\rm nonth}+4.2$
 and $L_{\rm pwn}=5 L_{\rm psr}^{\rm nonth}$,
 respectively.
The dash-dot lines in the {\em bottom} panel are lines of constant
PWN efficiency, $\eta_{\rm pwn}$. The error bars include the statistical
uncertainties (see Table 2) and the nominal 30\% distance uncertainties, except
for the Vela pulsar whose parallax has been measured
(Dodson et al.\ 2003).
 }
\end{figure}

It is also interesting to compare the PWN X-ray luminosities
with the magnetospheric (PL component) pulsar luminosities.
We see from Figure 7 that the former exceeds the latter for
9 out of the 10 objects in our sample (the only exception is the
Duck PWN and its pulsar B1757--24). Moreover, we see a clear
correlation between the two luminosities, which
 can be described as $\log L_{\rm pwn}=0.89\log L_{\rm psr}^{\rm nonth}+4.2$
or, more crudely,
\be
L_{\rm pwn} \approx 5 L_{\rm psr}^{\rm nonth},
\ee
in the 0.5--8 keV band.
Overall, the correlation is surprisingly strong, given the
 fact
that $L_{\rm psr}^{\rm nonth}$ was estimated assuming isotropic
pulsar emission
while the actual pulsar luminosity can be higher or lower than this
estimate.
The main outliers from this dependence
are the Vela, whose
pulsar is unusually underluminous (Pavlov et al.\ 2001a),
and the Duck/B1757--24,
for which separation of the pulsar and PWN emission is rather unreliable
even with the {\sl Chandra} resolution (Kaspi et al.\ 2001a).
On the other hand, B1800, for which $L_{\rm pwn}/L_{\rm psr}^{\rm nonth}\approx 4$,
is a typical PSR/PWN pair in this regard.
A natural reason for such a correlation is that both the magnetospheric
and PWN radiation are powered by the spin-down energy loss
(i.e. both $L_{\rm pwn}$ and $L_{\rm psr}^{\rm nonth}$ are
correlated with $\dot{E}$ and hence with each other).
However, the $L_{\rm pwn}$-$L_{\rm psr}^{\rm nonth}$ correlation
shows a smaller scatter than the $L_{\rm pwn}$-$\dot{E}$ correlation,
despite the quite different nature
of the magnetospheric and PWN
X-ray emission.
This may suggest that the $L_{\rm pwn}$-$\dot{E}$
 correlation is significantly distorted
by the distance errors
(in contrast to the $L_{\rm pwn}$-$L_{\rm psr}^{\rm nonth}$ correlation
that does not depend on distance)
  or, more likely, that
 there are some
other factors (e.g., the angle between the spin and magnetic axes) that
similarly affect the magnetospheric and PWN emission.

The spectral slope of the B1800 PWN is somewhat uncertain because
of the small number of counts and the dependence of
the fitting parameter $\Gamma_{\rm pwn}$ on
the poorly known $n_{\rm H}$ (see Figs.\ 3 and 4).
For a plausible $n_{\rm H,22}=1.4$, the photon index,
 $\Gamma_{\rm pwn} \simeq 1.6\pm 0.3$
for the entire PWN, is typical for the whole sample of PWNe
observed in X-rays\footnote{We should caution that the photon index
generally grows with increasing distance from the pulsar due the
synchrotron cooling. Therefore, the $\Gamma_{\rm pwn}$ value
depends on the extraction area chosen, and it may be systematically
smaller for more distant and fainter PWNe, where the outer regions with
softer emission are too faint to include them in the spectral analysis.
This observational bias must be taken into account when comparing PWN spectra.},
and it is substantially larger than
$\Gamma_{\rm pwn} =
2.36 -2.1
\dot{E}_{36}^{-1/2}=0.94$
(where $\dot{E}_{36} = \dot{E}/10^{36}\, {\rm ergs\, s}^{-1}$)
predicted by the $\Gamma_{\rm pwn}$-$\dot{E}$ correlation
found by Gotthelf (2003) for a sample of more energetic
pulsars. On the other hand, a lower assumed $n_{\rm H,22}=0.7$,
which cannot be ruled out based on the data available,
results in
an unusually hard PWN spectrum,
$\Gamma_{\rm pwn}\simeq 1.0\pm 0.2$.
For the broad range of $n_{\rm H}$ considered,
$\Gamma_{\rm pwn}$ remains similar to those of the other PWNe
in Table 2 except for the Mouse and the Duck, whose
softer spectra ($\Gamma_{\rm pwn}\sim 2.0$--2.5)
 can possibly be explained by
a stronger effect of  synchrotron cooling
in the long PWN tails.

The spectrum of the B1800 pulsar is even more uncertain
than that of its PWN.
As we have mentioned in \S2.2.2, the one-component PL model does not
fit the pulsar spectrum
for
$n_{\rm H,22}=1.4$, while the PL component
of the PL+BB fit gives $\Gamma_{\rm psr} \simeq 1.4\pm 0.6$,
similar $\Gamma_{\rm pwn}$ of B1800
 and to the spectral slopes of many other pulsars.
This slope is much softer than
$\Gamma_{\rm psr} =
2.1-2.9 \dot{E}_{36}^{-1/2}=0.1$
predicted
by the Gotthelf's (2003) correlation.
Lowering the assumed $n_{\rm H}$ by a factor of 2
makes a single-component PL fit acceptable, resulting in a rather soft
spectrum, $\Gamma_{\rm psr}\simeq 2.0\pm 0.4$.
The PL+BB fit at this smaller $n_{\rm H}$ yields a harder
PL component,
with $\Gamma_{\rm psr} \approx 1.2$ (see Fig.\ 5), which is close to
the $\Gamma_{\rm pwn}$ obtained for this $n_{\rm H}$,
but still considerably softer the Gotthelf's (2003) prediction.
Overall, we can only state that $\Gamma_{\rm psr}$ for B1800 is
within the range of values observed for the whole sample of radio
pulsars detected in X-rays, including the Vela-like pulsars.

The possible thermal component of the PL+BB fit to the B1800 pulsar
spectrum is very poorly constrained, not only because of the poor
statistics but also because the soft thermal radiation
 is strongly absorbed by the ISM. The BB temperature,
$T \sim 1$--3 MK, and bolometric luminosity, $L_{\rm psr}^{\rm bol}
\sim 10^{31}$--$10^{33}$ ergs s$^{-1}$,
are similar to those found from the PL+BB fits
for other Vela-like pulsars.
Such temperatures
are somewhat higher than the bulk NS surface temperatures
 predicted by the standard NS cooling
models (e.g., Yakovlev \& Pethick 2004), and the corresponding
emitting areas are smaller than the NS surface area.
However, the actual spectrum of the NS thermal radiation can differ
substantially from the BB model. In particular, fitting the spectra
with the hydrogen atmosphere
models (Pavlov et al.\ 1995) yields lower effective temperatures
and larger emitting areas, with not so strongly different bolometric
luminosities (see Pavlov et al.\ 2001a for the specific
example of the Vela pulsar). Unfortunately, the quality of the data
do not warrant fits with more complicated atmosphere models,
and even the estimate for the bolometric luminosity is too
uncertain to make a useful comparison with the NS cooling models.

\begin{figure}[t]
 \centering
\includegraphics[width=3.2in,angle=0]{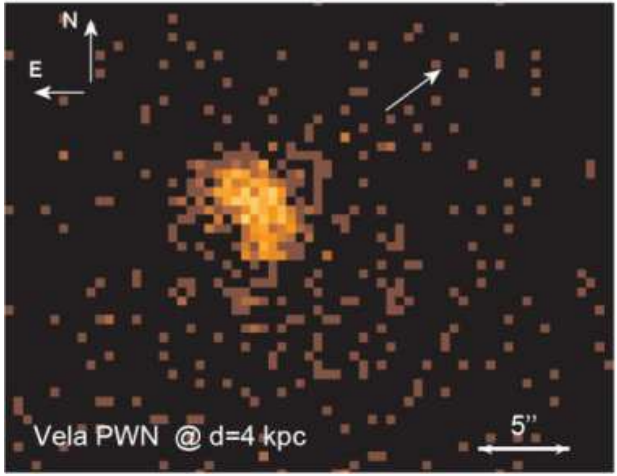}
 \caption{ {\sl Chandra} ACIS image of the Vela PWN in the 1--8 keV band
 as it would be seen at the distance of 4 kpc in a 30 ks exposure.  The resolution is degraded
 by a factor 13 to simulate the Vela PWN appearance
 at the distance of 4 kpc. The arrow shows the direction of the Vela pulsar proper motion.
}
\end{figure}

\subsection{
Inferences from the PWN morphology
}

The distribution of X-ray counts
 in the immediate vicinity of the B1800 pulsar shows that the X-ray PWN is
 elongated in the direction perpendicular to the pulsar's proper
 motion.
By analogy with young, bright PWNe
(such as the Crab and Vela),
 the observed PWN shape
can be
 interpreted as a torus with a symmetry axis along
the southwest-northeast
 direction
 (see Fig.\ 1). The torus appears to be seen nearly edge-on,
 implying that the
pulsar's spin axis is close to
the plane of the sky. We find no evidence for jets along the
 symmetry axis. This, however, should not be surprising because
  such jets can be significantly fainter
 than the torus
(for instance, the Vela PWN
outer jets would not be seen in a 30 ks exposure if
the Vela
 PWN were located at a 4 kpc distance
while the inner jets would not be resolved; see Fig.\ 8).

In the B1800 PWN,
the torus symmetry axis, which presumably
coincides with the pulsar spin axis,
is approximately aligned (within $\approx 10^{\circ}$) with the
direction of the pulsar's proper motion. This
 strengthens the evidence for the alignment between the pulsar spin
and velocity vectors,
based
 on a very limited sample of pulsars so far
(e.g., Ng \& Romani 2004; Johnston et al.\ 2005),
which constrains the physics
of SN explosions.
It also implies that
the total (three-dimensional) speed
of the pulsar is close to the measured transverse
velocity in the plane of the sky, $v_\perp =
(365\pm 30) d_4$ km s$^{-1}$
(Brisken et al.\ 2006).

Toroidal (or arc-like) X-ray structures are often
  found around young pulsars,
the Vela and the Crab
being the most famous examples.
Out of the ten Vela-like pulsars observed with
   {\sl Chandra} (see Table
2 and Fig.\ 9), three (Vela, B1706--44, and J2021+3651) clearly show
arc-like (or toroidal) structures and jets in X-rays.
 On the other hand, the PWNe around J1747--2958 (the Mouse) and
B1757--24 (the Duck)
exhibit different
X-ray morphologies, with prominent tails behind the moving
pulsars,
indicative of bow-shock
PWNe.
 For the remaining
four pulsars (B1823$-$13, J1016$-$5857, B1046$-$58, and J1811$-$1925), the
existing X-ray data
do not allow one
to firmly claim the existence of either a tail or a torus based on
X-ray data alone.

\begin{figure*}[t]
 \centering
\includegraphics[width=4.4in,angle=0]{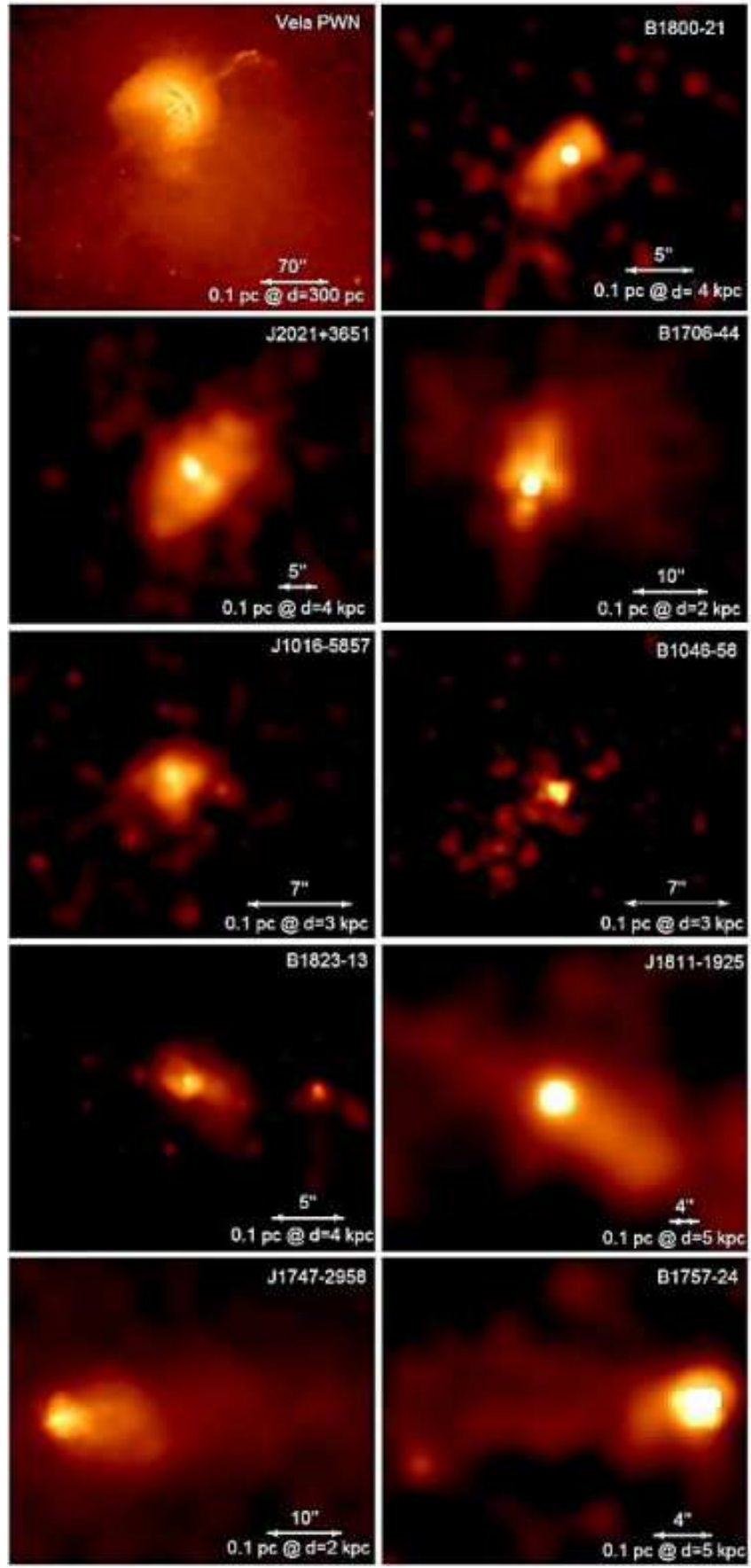}
 \caption{ {\sl Chandra} ACIS images of PWNe around Vela-like pulsars
(see Table 2).}
\end{figure*}

Despite the relatively high
speed of B1800,
the PWN shows no
evidence of
bow-shock morphology. This may look surprising since,
for instance, the Mouse PWN,
whose pulsar has almost the same age and spin-down power as B1800,
clearly shows such
morphology, although the speed
of its pulsar,
inferred from the
bow-shock modeling,
is only a factor of 1.6 higher (Gaensler et al.\ 2004)\footnote{We should
note that the proper motion of PSR J1747--2958 has not been measured.
The Mach number ${\mathcal M}\equiv v/c_s=60$
was obtained by Gaensler at al.\ (2004)
assuming an ambient pressure $p_{\rm amb}=3.3\times 10^{-13}$
ergs cm$^{-3}$, and the pulsar speed $v=600$ km s$^{-1}$ was estimated
assuming that the ambient medium is a warm ISM with the sound speed
$c_s\approx 100$ km s$^{-1}$. Both ${\mathcal M}$ and $v$ would be smaller
if the ambient pressure is higher than assumed, as one can expect for such
a young pulsar ($\tau = 26$ kyr), which is possibly at outskirts of
its SNR. For instance,
${\mathcal M}=25
(d/5\,{\rm kpc})^{-1}$ for $p_{\rm amb} =
2\times 10^{-12}\,{\rm ergs}\,{\rm cm}^{-3}$, and $v=355 n^{-1/2} (d/5\,{\rm kpc})^{-1}$ km s$^{-1}$, close
to the transverse velocity of B1800.}.
 Furthermore, for the Duck PWN,
recent radio-interferometric
measurements
put an upper limit of $340\,(d/5\,{\rm kpc})$ km s$^{-1}$ on the
transverse pulsar speed (Blazek et al.\ 2006). Nevertheless, the
Duck shows a distinct trail in both radio and X-rays
(Frail \& Kulkarni 1991; Kaspi et al.\ 2001a), which is a
signature of a ram-pressure confined, supersonically moving PWN.
Thus, the different appearance of the B1800
PWN as compared to the Duck and the Mouse  should be attributed to
 differences in the properties of ambient medium rather than to
different pulsar velocities.

Assuming
that the lack of bow-shock morphology in B1800 implies a subsonic
pulsar velocity, we obtain
a lower limit on the temperature of the
ambient medium,
$T\gtrsim
10 d_4^2$ MK.
Such
a high temperature can only be expected inside a
young SNR
(e.g., temperatures of up to
14 MK have been
measured in the vicinity of the Vela pulsar, although typical
temperatures in the Vela SNR are
1--3 MK;
Lu \& Aschenbach
  2000).
This means that the radius of the putative host SNR is substantially
larger than $v_\perp \tau \approx 6\, d_4$ pc ($R>5'$ in angular scale;
for comparison, the radius of the Vela SNR, $\sim 4^\circ$ at 300 pc,
would be $\sim 18'$ at $d=4$ kpc).
However, our analysis of the ACIS image
showed no evidence of
an SNR (see \S3.1), neither the putative host SNR of B1800 nor
the
G8.7--0.1 remnant onto which B1800 is projected.
If the X-ray spectrum and luminosity of the undected SNR are similar
to those of the Vela SNR, the nondetection is not surprising
because of the much larger $n_{\rm H}$ for B1800.
For instance, using
the PIMMS tool\footnote{See http://heasarc.gsfc.nasa.gov/Tools/w3pimms.html.},
we estimate that the average unabsorbed intensity from the Vela SNR,
$I \sim 1.5\times 10^{-13}$ ergs cm$^{-2}$ s$^{-1}$ arcmin$^{-2}$
in the 0.1--2.4 keV band (Lu \& Aschenbach 2000), corresponds to
the on-axis ACIS-S3 surface
brightness of $<0.3$ counts ks$^{-1}$ arcmin$^{-2}$
in the 0.5--7 keV band,
for the Raymond-Smith thermal plasma emission models with $T<3$ MK
and $n_{\rm H,22}=1.4$, i.e. at least a factor of 13 below the observed
upper limit.
Even the hot thermal component of the Vela SNR, such as observed in
the vicinity of the Vela pulsar, could hardly be detected in our observation
if Vela were at the location of B1800.
For instance, $I\sim 3\times 10^{-14}$ ergs
cm$^{-2}$ s$^{-1}$ arcmin$^{-2}$ in the 0.9--2.4 band, as estimated
from Figures 15 and 9 of Lu \& Aschenbach (2000), corresponds to
the ACIS-S3 surface brightness of only 1.2 counts ks$^{-1}$ arcmin$^{-2}$,
for $T=10$ MK
and $n_{\rm H,22}=1.4$.
On the other hand, Finley \& \"{O}gelman (1994) did detect
extended X-ray emission $\sim 30'$ northeast of the pulsar
(this region was out of the {\sl Chandra} ACIS field of view)
and interpreted it as due to a shock in the G8.7--0.1 SNR,
reflected from a nearby molecular cloud.
Fitting the spectrum of that emission with the Raymond-Smith
model, they found $T=4$--8 MK and $n_{\rm H,22}=1.2$--1.4.
Based on the currently availble data, we cannot conclude whether
this emission is associated with the putative host SNR of B1800
or with G8.7--0.1.
In the latter case, the G8.7--0.1 SNR should be at a distance
similar  to that of the
host SNR of B1800 (e.g., $\sim$1 kpc foreground)
 as follows from the similar $n_{\rm H}$ values.

The shape and size of the B1800 PWN can be used to obtain additional
constraints on the properties of the ambient medium, such as pressure,
density, and magnetic field.
For instance, the ambient pressure confining the PWN should be
a sizeable fraction of the pulsar wind pressure at the termination
shock,
$p_{\rm amb} = f_p p_s$, where $f_p\lesssim 1$,
\be
p_s = \dot{E} f_\Omega (4\pi c r_s^2)^{-1}\, ,
\ee
 $r_s$ is the termination
shock radius in the equatorial plane, and the factor $f_\Omega$ takes
into account anisotropy of the pulsar wind (for instance, $f_\Omega\approx 3/2$
for the often assumed dependence of the wind flux, $F_w\propto
\sin^2\theta$, on rotational colatitude $\theta$; e.g., Bogovalov
\& Khangoulyan 2002).
In well-resolved bright PWNe, such as the Crab and Vela,
we see the ``inner ring'', commonly associated with the pulsar
wind termination shock in the equatorial outflow.
We cannot resolve such a ring in the B1800 image, but,
by analogy with the bright, well-resolved PWNe, we
can assume its radius to be about one half of the radial
extent of the X-ray bright torus in equatorial plane.
Based on the
surface brightness profile shown in Figure 1 (bottom right panel), we
take
  $r_s\approx 1\farcs5
\approx0.03d_4$ pc
as a reasonable estimate,
which is close to the $r_s$ values
measured in other Vela-like X-ray PWNe
    with clearly visible toroidal structures
(including Vela, B1706--44, and J2021+3651; see Table 2).
Then, the ambient pressure can be estimated as
\be
p_{\rm amb}\sim 7.2\times 10^{-10} f_p f_\Omega d_{4}^{-2}\,\, {\rm ergs\,\, cm}^{-3}.
\ee
Similar to many other PWNe around young pulsars (see Fig.\ 10),
the estimated
  ambient pressure is
much higher than
the typical unperturbed ISM pressure, $\sim 10^{-12}$
ergs
cm$^{-3}$, in agreement with our inference that B1800 has not
left its SNR.

Using a similar approach,
we estimated
the pulsar wind pressure at the termination
shock for 13 other young PWNe,
including the Mouse and the Duck for which
$r_s$ is the estimated separation between the pulsar
and the bow-shock head, and
plotted $p_s$ versus
pulsar spin-down age
in Figure 10, assuming $f_\Omega =1$.
Even with allowance for large uncertainties,
the pressure does not follow the
dependence $p_{\rm amb}\propto \tau^{-6/5}$ expected
for SNRs in Sedov stage, nor any other smooth dependence.
Particularly surprising is the low pressure,
$p_s \sim 4\times 10^{-11} f_\Omega (d/4.4\,{\rm kpc})^{-2}$ ergs cm$^{-3}$,
 for the very
young ($\tau\simeq 1.6$ kyr) PSR B1509--58, which suggests a very
unusual SNR for this object, in agreement with X-ray observations
(Gaensler et al.\ 2002).
It also hints that PSR B1509--58 moves supersonically
(which would require a velocity above
$50\, (d/4.4\,{\rm kpc})^{-1} n^{-1/2}$ km s$^{-1}$,
where $n=\rho/m_H$),
so that
the arc(s) northwest of the pulsar and the bright, long
``jet'' southeast of the pulsar are elements of a bow-shock PWN
rather than a ``torus-jet'' PWN.
We should also note that the
spin-down age of a pulsar can
differ substantially from its true age, which can contribute to
 the lack of a clear $p_s$-$\tau$ correlation
in Figure 10. For instance, $\tau=620$ kyr for PSR J0538+2817
is a factor of 6 larger than the estimated age of the G180.1--1.7 SNR,
presumably associated with this pulsar (e.g., Romani \& Ng 2003).

Interestingly, the pulsar wind pressure at the head of bow-shock
PWNe is similar to that at the termination shock radius of torus-like
PWNe, despite the different confining mechanisms. The reason for this
coincidence is that, for typical pulsar speeds,
 the ram pressure caused by the pulsar motion
in the ISM,
$p_{\rm ram}=\rho v^2 = 1.5\times 10^{-9} n (v/300\,{\rm km\,s}^{-1})^2$
ergs cm$^{-3}$, is of the same order of magnitude as the ambient
pressure inside an SNR confining a torus-like PWN.

Using the above estimate for the ambient pressure confining B1800
and assuming that the ram pressure due to the pulsar's
motion,
\be
p_{\rm ram}
\approx 2.2\times 10^{-9} n d_4^2\,\, {\rm ergs\,\,
cm}^{-3},
\ee
 is lower than $p_{\rm amb}$ (as expected for the
subsonic motion), we obtain an upper limit on the density of the
ambient medium, $n < 0.3 f_p f_\Omega d_4^{-4}$ cm$^{-3}$,
or $\rho < 5\times 10^{-25} f_p f_\Omega d_4^{-4}$ g cm$^{-3}$.

The measured termination shock radius can also be used to
estimate the magnetic field
inside the X-ray PWN.
According to Kennel \& Coroniti (1984), the particle pressure
immediately downstream of a strong termination shock
perpendicular to the wind (i.e., close to the equatorial
plane in our case) is
$2p_s/3$, so that the internal energy is $2p_s$.
If $\epsilon_B$ is the fraction of the internal energy in the
magnetic field, the field can be estimated as
$B=(16 \pi \epsilon_B p_s)^{1/2}$, or
\be
B=\left(\frac{4\epsilon_B \dot{E} f_\Omega}{r_s^2}\right)^{1/2} \approx 190\, \epsilon_B^{1/2} f_\Omega^{1/2} d_4^{-1}\,\mu{\rm G}.
\ee
The fraction $\epsilon_B$ varies along the post-shock flow,
and it depends on the magnetization and angular distribution of the pre-shock
pulsar wind, as well as on the properties of the termination shock.
For instance, in the Kennel \& Coroniti (1984) model of spherical
shock in an isotropic wind with a small pre-shock magnetization
parameter $\sigma$, the magnetic energy fraction is
 $\epsilon_B \approx 9\sigma/4$ immediately downstream of the shock,
and it grows with increasing radius up to $\epsilon_B \sim 1$
at $r\approx r_s(3\sigma)^{-1/2}$. In a more realistic
equatorial outflow
geometry, Komissarov \& Lyubarsky (2004) found a highly nonuniform
distribution of $\epsilon_B$ in the postshock flow, with maximum
values close to unity. Given the uncertainty of the flow pattern in
B1800, we can only crudely estimate $B\sim 100\,\mu$G for a typical
magnetic field in its PWN, similar to the Vela and other
Vela-like PWNe (e.g., Gonzalez et al.\ 2006).

A crude estimate of the synchrotron cooling time for
electrons emitting at an energy $E$ (in keV)
is $\tau_{\rm syn}
\approx 36 B_{-4}^{-3/2} E^{-1/2}$ yrs, where $B_{-4}=B/100\,\mu$G.
During the time $\tau_{\rm syn }$, the pulsar travels
the projected distance, $v_{\perp}\tau_{\rm syn }\approx
0.013 d_4 E^{-1/2} B_{-4}^{-3/2}$ pc,
corresponding to the angular distance of
$0\farcs7 E^{-1/2} B_{-4}^{-3/2}$,
considerably smaller than the size of the compact X-ray PWN.
For a typical shocked wind flow velocity of $\sim 0.3 c$,
the electrons travel a distance of $3.3 E^{-1/2} B_{-4}^{-3/2}$ pc
during the synchrotron cooling time, corresponding to an angular
distance of $2\farcm8\, d_4 E^{-1/2} B_{-4}^{-3/2}$, which substantially
exceeds the observed extent of the X-ray PWN. We caution, however,
that the flow streamlines are curved for an anisotropic pulsar
wind (see, e.g., Komissarov \& Lyubarsky 2004), so that the actual
PWN size can be smaller than this distance. In addition,
the synchrotron
surface brightness decreases with the distance from the pulsar (see
e.g., the Vela PWN image in Fig.\ 9) rendering fainter emission
undetectable during the relatively short ACIS exposure. This could
also explain the apparent lack of spectral softening (\S2.2.1) due
to the synchrotron burn-off, which should become more significant
further away from the pulsar. A deeper exposure can reveal the
expected faint
extension of the X-ray torus beyond the $5''$ radius if the equatorial
flow remains sufficiently well collimated.

The compact inner PWN of B1800, which we interpret as a torus
seen edge-on, shows some asymmetry with respect to the spin axis:
the projected torus is more elongated towards the southeast,
and its southeast edge is more diffuse and fainter than the
northwest edge (see Fig.\ 1). A possible explanation for this
asymmetry is a pressure
gradient in the ambient SNR matter.
As $r_s \propto p_{\rm amb}^{-1}$, the pressure
gradient leads to an azimuthal
dependence of the termination shock radius in the equatorial
plane,
such that the circular ring turns into an ellipse-like structure.
Since the surface of such an asymmetric shock is not perpendicular to
the unshocked pulsar wind even in the equatorial plane,
the post-shock flow
   is deflected
toward the most distant (southeast) part of the shock, where the magnetic
field and the synchrotron emissivity are lower.
 To produce the observed asymmetry,
the pressure difference should be a sizable fraction of the mean
ambient pressure,
$\Delta p \sim {\rm a\,\,\, few}\, \times 10^{-10}$ ergs cm$^{-3}$.
Such a difference should cause a local wind in the ambient
medium, possibly on a parsec scale,
 with a typical velocity
$v_{\rm amb} \sim (\Delta p/\rho)^{1/2}
\sim 240 (\Delta p/3\times 10^{-10}\,{\rm ergs\,\, cm}^{-3})^{1/2}
(\rho/5\times 10^{-25}\,{\rm g\,\, cm}^{-3})^{-1/2}$ km s$^{-1}$,
comparable to the pulsar's velocity. This wind,
blowing, e.g., from the west with a speed comparable to
that of the pulsar,
could also explain the asymmetry of the fainter PWN emission,
extended  toward southeast-south
 (see Fig.\ 2). Similar
asymmetry
is seen in
 the brighter nearby Vela PWN (see Fig.\ 9),
which has also been attributed to a local SNR wind
(Pavlov et al.\ 2003).
Such pressure gradients and accompanying winds could also be responsible
for the asymmetric shape of other PWNe, such as B1823--13 (see Fig.\ 9).

\begin{figure}[]
 \centering
\includegraphics[width=2.7in,angle=90]{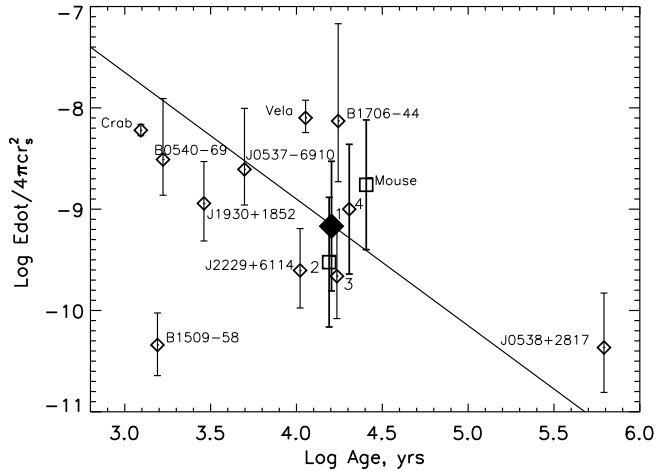}
\caption{
Pulsar wind pressure at termination shock versus
pulsar
spin-down age for a sample of 14 pulsars,
assuming an isotropic pulsar wind.
 The diamonds mark PWNe showing
torus-like structure (B1800 is shown by the large filled diamond),
while the two PWNe with a clear bow-shock structure are shown by
squares.
The pulsars B1800--21, B1757--24 (the Duck), J2021+3651, and B1046--58 are
 denoted by the numbers 1, 2, 3 and 4, respectively.
To calculate the shock radius $r_s$, we used the distance $d$ in
the sixth column of Table 2.
The straight line, arbitrarily drawn through the B1800 point,
 shows the dependence $p\propto \tau^{-6/5}$
expected for an SNR in Sedov stage.
 }
\end{figure}

\section{Conclusions}

We have detected the X-ray emission from PSR B1800$-$21 and its synchrotron
nebula,
elongated perpendicular to the direction of the pulsar's proper motion.
The shape of the observed PWN suggests that, similar to many other young
pulsars, B1800 is surrounded by a compact X-ray torus
around an equatorial termination shock in the pulsar
wind.
The symmetry
axis  of the torus (i.e.\ the pulsar's spin axis) is approximately
aligned with the direction of the proper
motion. The compact nebula is embedded into a fainter, more extended emission,
which shows substantial asymmetry, possibly caused by
a pressure gradient in the ambient medium.
Despite a rather high
velocity of B1800, we
find no evidence of a bow-shock X-ray nebula.
This implies that the pulsar is still moving in a hot SNR medium,
but we do not see the SNR in the {\sl Chandra} image.
In terms of its X-ray efficiency and spectrum, the B1800 PWN is very
similar to PWNe around other pulsars of similar ages and spin-down powers,
 subsonically moving within their SNRs. Bow-shock PWNe associated with
young, supersonically moving pulsars show higher X-ray efficiencies
and softer spectra.

The spectrum of the B1800 pulsar
 can be described by a two-component BB+PL model. For reasonable
 $n_{\rm H}$ values, the parameters of this
model
 and the corresponding component luminosities resemble those
 of other Vela-like pulsars.

 A deeper {\sl Chandra} observation would
allow one to investigate
the morphology of a fainter outer PWN and perhaps reveal the jets
along the torus symmetry axis. To look for the putative host SNR
and investigate the PWN emission on a larger spatial scale,
including its possible connection with the TeV source HESS J1804-216,
this field should be observed with {\sl XMM-Newton}.

\acknowledgements
This work was partially supported by NASA grants NAG5-10865
and NAS8-01128 and {\sl Chandra} awards AR5-606X and SV4-74018.

\begin{table*}[]
\caption[]{Properties of Vela-like pulsars observed with {\sl
Chandra}} \vspace{-0.5cm}
\begin{center}
\begin{tabular}{lccccccccccccccc}
\tableline\tableline PSR &  $P$  & $\tau$ &
$\dot{E}_{36}$\tablenotemark{a} & $d_{\rm DM}$\tablenotemark{b}  & $d$\tablenotemark{c} & $n_{\rm H,22}$\tablenotemark{d} & $L_{\rm pwn,32}\tablenotemark{e}$ & $\Gamma_{\rm pwn}$ & $L_{\rm psr, 32}^{\rm nonth}\tablenotemark{f}$ & $\Gamma_{\rm psr}$ & $L^{\rm bol}_{\rm psr,32}$~\tablenotemark{g} & $l_{X}\tablenotemark{h}$ & $r_s\tablenotemark{j}$ &
$p_{s}\tablenotemark{k}$ & Ref.\tablenotemark{l} \\
\tableline
&  ms  & kyr &  & kpc  & kpc & & &  &  &  &   & pc & pc & $10^{-9}$ cgs & \\
\tableline
Vela             &       $~89$                     &      $11.3$             &       $6.9$   &  $0.24$  & $~~0.3$ &  0.02& $1.3\pm0.1$    & $1.4\pm0.1$         & $0.3\pm0.1$  & $2.0\pm0.3$         & $1.4\pm0.4$     & $0.5$ & 0.016 & 8.0 & 1\\
J1811--1925       &       $~65$                     &      $23.3$             &       $6.4$   &  ...     & 5   & 3.1& $101\pm20$     & $1.5\pm0.2$         & $70\pm2$        & $1.4\pm0.1$         &  ...                           & $1.4$     & ...   & ... & 2           \\
B1706--44         &       $102$                    &      $17.5$             &       $3.4$   &  $2.30$  & 2   & 0.5& $3.8\pm0.15$   & $1.8\pm0.1$        & $1.1\pm0.1$    & $1.7\pm0.2$       & $3.4\pm0.8$   & $0.4$  & 0.012   & $7.4$ & 3\\
J2021+3651       &       $104$                    &      $17.2$             &       $3.4$   &  $12.4$  & 4   & 0.7 & $12\pm2$       & $1.7\pm0.3$ & $1.8\pm0.3$     & $1.0^{+0.6}_{-0.3}$ & $9^{+5}_{-4}$                   & $0.5$       & 0.068   & 0.22  & 4          \\
 B1823--13         &       $101$                    &      $21.4$             &       $2.8$   &  $3.93$  & 4   & 1.7 & $3.9\pm0.4$   & $1.7\pm0.4$         & $1.6\pm0.4$    & $2.2\pm0.4$                       & ...                       & $0.2$     & ...   & ...    & 5     \\
J1016--5857       &       $107$                    &      $21.0$             &       $2.6$   &  $8.00$  & 3   & [1.2] & $2.0\pm0.5$    & $1.5\pm0.2$       & $0.5\pm0.2$    & $1.5\pm 0.4$       & $5^{+5}_{-3}$     & $0.1$    & ...   & ...     & 6       \\
B1757--24         &       125               &      15.5      &       2.6   &  5.22  & 5 & 4.4  & $16\pm5$       &$2.5\pm0.3$         & $17\pm2$        & $1.9\pm0.3$         & ...                       & $0.5$     & 0.05 & 0.3  & 7\\
J1747--2958       &       $~99$                     &      $25.5$             &       $2.5$   &  $2.01$  & 5   & 3.0 & $500\pm60$      & $2.0\pm0.2$         & $55\pm3$     & $1.6\pm0.1$                  & ...                      & $0.6$     & 0.02 & 1.7  & 8 \\
{\bf B1800--21}      &       134                    &      15.8             &       2.2   &  3.88  & 4   & 1.4 & $1.6\pm0.2$   & $1.6\pm0.3$           & $0.4\pm0.1$   & $1.4\pm0.6$           & $3^{+5}_{-2}$               & 0.2          & 0.02 & 0.7 & ... \\
B1046--58         &       124                    &      20.3         &       2.0   &  2.73  & 3   & [0.4] & $0.66\pm0.06$  & $1.0\pm0.2$         & $0.26\pm0.04$   & $1.5\pm0.3$                          & ...              & $0.2$    & 0.024 & 1.0 & 9\\
\tableline
\end{tabular}
\end{center}
\tablecomments{To ensure uniformity of analysis, we use the results of our own analysis of the {\sl Chandra} data.
 These results may differ from the published ones because of different extraction
regions used, different $n_{\rm H}$ assumed, etc. The uncertainties of the luminosities and photon
indices include
the statistical uncertainties from spectral fitting but not the systematic uncertainties which can be substantial in some cases and are
 difficult to calculate.
}
\tablenotetext{a}{Spin-down power in units of $10^{36}$ ergs s$^{-1}$.}
\tablenotetext{b}{Dispersion measure distance from the Cordes \&
Lazio (2002) model. }
\tablenotetext{c}{Our best guess for a plausible
distance to the pulsar, used to scale
all the distance-dependent
parameters in this table. For three pulsars,
J2021+3651, J1016--5857, and J1747--2958, the adopted distances differ substantially
 from $d_{\rm DM}$.
 In the two cases with the
largest DM distances
our distance estimates are
based on the measured $n_H$ and the Galactic HI column.
For the detailed discussion of the distance to J1747--2958, see
 Gaensler et al.\ (2004).
}
\tablenotetext{d}{
The hydrogen column density is obtained from the spectral fits to the
PWN spectra, except for J1016-5857 and B1046-58 for which the
small number of PWN counts precludes a reliable $n_{\rm H}$ measurement. In these two cases
the $n_{\rm H}$ values (shown in square brackets) are estimated  from the pulsar's DM (assuming 10\% ISM ionization).
\tablenotetext{e}{Unabsorbed PWN luminosity in the 0.5--8 keV band, in units of $10^{32}$ ergs
s$^{-1}$. }
For the Vela PWN we quote
the luminosity of the inner compact PWN restricted to the ``arcs''
region (i.e. within $\approx25''$ from the pulsar).}
\tablenotetext{f}{Nonthermal luminosity of the pulsar in the 0.5--8 keV band, in units of $10^{32}$
ergs s$^{-1}$.
In the cases when the spectrum
is fitted with
the BB+PL model, it is the luminosity of
the PL component only. }
\tablenotetext{g}{Thermal bolometric luminosity of
the BB component from the PL+BB fit in units of $10^{32}$ ergs s$^{-1}$.}
\tablenotetext{h}{The largest linear extent of the observed X-ray
emission from a PWN.}
\tablenotetext{j}{Estimated stand-off
distance from the pulsar to the termination shock.}
\tablenotetext{k}{Estimated pressure at the
termination shock assuming $f_{\Omega}=1$ (see text).}
\tablenotetext{l}{References to
papers where the corresponding {\sl Chandra} data have been
analyzed. -- (1) Pavlov et al.\ 2001ab; (2) Roberts et al.\ 2003; (3) Romani et al.\ 2005; (4) Hessels et al.\ 2004; (5); Teter et al. 2003; (6) Camilo et al.\ 2004; (7) Kaspi et al.\ 2001a; (8) Gaensler et al.\ 2004;
(9) Gonzalez et al.\ 2006.}

\end{table*}

\end{document}